\documentclass[conference,compsoc]{IEEEtran}
\IEEEoverridecommandlockouts

\usepackage{cite}
\usepackage{amsmath,amssymb,amsfonts}
\usepackage{algorithmic}
\usepackage{graphicx}
\usepackage{textcomp}
\usepackage{xcolor}
\def\BibTeX{{\rm B\kern-.05em{\sc i\kern-.025em b}\kern-.08em
    T\kern-.1667em\lower.7ex\hbox{E}\kern-.125emX}}

\usepackage{subcaption}
\usepackage{booktabs}
\usepackage{caption}
\usepackage{multirow}
\usepackage{tcolorbox}
\usepackage{xspace}
\usepackage{hyperref}
\usepackage{array,multirow,graphicx}
\usepackage{listings}
\definecolor{dkgreen}{rgb}{0,0.5,0}
\definecolor{lessdkgreen}{rgb}{0,0.6,0}
\definecolor{dkred}{rgb}{0.5,0,0}
\definecolor{gray}{rgb}{0.5,0.5,0.5}

\lstdefinestyle{cstyle}{
language=c,
basicstyle=\ttfamily\bfseries\scriptsize,
  morekeywords={virtualinvoke},
  keywordstyle=\color{blue},
  ndkeywordstyle=\color{red},
  commentstyle=\color{dkred},
  stringstyle=\color{dkgreen},
  numbers=left,
  breaklines=true,
  numberstyle=\ttfamily\footnotesize\color{gray},
  stepnumber=1,
  numbersep=2pt,
  numberstyle=\tiny,
  backgroundcolor=\color{white},
  tabsize=4,
  showspaces=false,
  showstringspaces=false,
  xleftmargin=.23in,
  captionpos=b,
  escapeinside={$}{$},
  print
}
\lstdefinestyle{cinlinestyle}{
language=c,
basicstyle=\ttfamily\bfseries\footnotesize,
  morekeywords={virtualinvoke},
  keywordstyle=\color{blue},
  ndkeywordstyle=\color{red},
  commentstyle=\color{dkred},
  stringstyle=\color{dkgreen},
  escapeinside={$}{$},
  print
}

\newcommand\cinline[1]{{\lstinline[style=cinlinestyle]@#1@}}

\newcommand{\realtype}{\textsc{realtype}\xspace}
\newcommand{\idioms}{\textsc{idioms}\xspace}
\newcommand{\exebench}{\textsc{exebench}\xspace}

\begin{document}

\title{Idioms: Neural Decompilation With Joint Code and Type Definition Prediction
}

\author{\IEEEauthorblockN{Luke Dramko}
\IEEEauthorblockA{\textit{School of Computer Science} \\
\textit{Carnegie Mellon University}\\
lukedram@cs.cmu.edu}
\and
\IEEEauthorblockN{Claire Le Goues}
\IEEEauthorblockA{\textit{School of Computer Science} \\
\textit{Carnegie Mellon University}\\
clegoues@cs.cmu.edu}
\and
\IEEEauthorblockN{Edward J. Schwartz}
\IEEEauthorblockA{\textit{Software Engineering Institute} \\
\textit{Carnegie Mellon University}\\
eschwartz@cert.org}
}

\maketitle

\begin{abstract}
Decompilers are important tools for reverse engineers that help them analyze software at a higher level of abstraction than assembly code.
Unfortunately, because compilation is lossy, deterministic decompilers produce code that is missing many of the details that make source code readable in the first place, like variable names and types.
Neural decompilers, on the other hand, offer the ability to statistically fill in these details.
Existing work in neural decompilation, however, suffers from substantial limitations that preclude its use on real code, such as the inability to define composite types, which is essential to fully specify function semantics.
In this work, we introduce a new dataset, \realtype, that includes substantially more complicated and realistic types than existing neural decompilation benchmarks, and \idioms, a new neural decompilation approach to finetune any LLM into a neural decompiler capable of generating the appropriate user-defined type definitions alongside the decompiled code.
We show that our approach yields state-of-the-art results in neural decompilation. On the most challenging existing benchmark---\exebench---our model achieves 54.4\% accuracy vs.\ 46.3\% for LLM4Decompile and 37.5\% for Nova; on \realtype, our model performs at least 95\% better.

\end{abstract}


\section{Introduction}
\label{sec:introduction}

Decompilation---the reconstruction of a source code representation from an
executable program---is useful for a variety of security tasks, including malware analysis,
vulnerability research, and fixing legacy software when the original source
code is unavailable.
Unfortunately, because the compilation process loses many programmer-oriented abstractions, such as variable names, types, and comments,
the code produced by traditional deterministic compilers is often difficult to read and understand.

To address these problems, researchers have been applying machine learning,
which offers the possibility to guess or predict such missing abstractions
\emph{statistically} based on the surrounding context.
Some work restricts itself to recover predefined abstractions, such as variable names~\cite{dire,direct,varbert}, function names~\cite{nero,symlm,kim2023transformer}, variable types~\cite{lehmann2022finding,osprey,tygr}, or several abstractions at once~\cite{dirty,hext5,resym}. 
While promising, there are numerous issues with decompiled code~\cite{dramko2024taxonomy} and maintaining a model for each one is challenging.

More recently, researchers have leveraged large language models (LLMs) to recover the original source code, rather than specific abstractions, which we call \emph{neural decompilation}~\cite{nova,degpt,llm4decompile}.
Neural decompilation is appealing because, in theory, it can statistically recover any type of abstraction that is missing or distorted.
In some cases, it can greatly outperform traditional, deterministic decompilers.

For example, Figure~\ref{fig:intro_original} shows a function that finds the index of an element in a hash table which uses robin-hood hashing for collision management.
Figure~\ref{fig:intro_decompiled} shows the same function, but after having been compiled and decompiled with the industry-standard Hex-Rays decompiler.
This decompiled code lacks meaningful names, while a pointer to the hash table is misrepresented as an \cinline{__int64}.
Figure~\ref{fig:intro_llm4decompile} shows the prediction of LLM4Decompile~\cite{llm4decompile}, a state-of-the-art neural decompiler.
Although LLM4Decompile did not recover meaningful names, it did predict that the function's first argument had a structure type, and converted the raw pointer arithmetic into more readable field accesses.

\begin{figure*}[!htb]
  \begin{subfigure}[t]{0.49\textwidth}
  \begin{lstlisting}[style=cstyle,basicstyle=\ttfamily\bfseries\scriptsize]
struct hash {
  int hash_size;
  int item_cnt;
  struct gap_array *data;
  int (*hash_make_key)(void *item);
  int (*cmp_item)(void *item1, void *item2);
}
struct gap_array {
  int len;
  void **array;
}
  
int hash_find_index(struct hash *h, void *item) {
    void *cnx;
    int index = hash_make_key(h, item);
    int cnt = 0;
    cnx = gap_get(h->data, index);
    while (cnx != NULL) {
        if (cnt++ > h->hash_size) return -1;
        if (!h->cmp_item(cnx, item)) break;
        index = hash_next_index(h, index);
        cnx = gap_get(h->data, index);
    }
    if (cnx == NULL) return -1;
    return index;
}
  \end{lstlisting}
  \caption{A function which finds the index of an element in a hash table where collisions are handled with robin-hood hashing.\label{fig:intro_original}}
  \end{subfigure}
  \hfill
  \begin{subfigure}[t]{0.49\textwidth}
  \begin{lstlisting}[style=cstyle,basicstyle=\ttfamily\bfseries\scriptsize]
  __int64 __fastcall func4(__int64 a1, __int64 a2) {
  int v2;          // eax
  __int64 result;  // rax
  int v4;          // [rsp+10h] [rbp-10h]
  unsigned int v5; // [rsp+14h] [rbp-Ch]
  __int64 i;       // [rsp+18h] [rbp-8h]
  v5 = func2(a1, a2);
  v4 = 0;
  for (i = func1(*(_QWORD *)(a1 + 8), v5); i;
       i = func1(*(_QWORD *)(a1 + 8), v5)) {
    v2 = v4++;
    if (v2 > *(_DWORD *)a1) return 0xFFFFFFFFLL;
    if (!(*(unsigned int(__fastcall **)(__int64, __int64))(a1 + 24))(i, a2))
      break;
    v5 = func3((_DWORD *)a1, v5);
  }
  if (i)
    result = v5;
  else
    result = 0xFFFFFFFFLL;
  return result;
}
  \end{lstlisting}
  \caption{Figure~\ref{fig:intro_original}, after being compiled, and then decompiled by Hex-Rays. The decompiled version is missing details like meaningful identifier names and types, which makes it harder to read.\label{fig:intro_decompiled}}
  \end{subfigure}
  
  \begin{subfigure}[b]{0.49\textwidth}
  \begin{lstlisting}[style=cstyle,basicstyle=\ttfamily\bfseries\scriptsize]
int FUN_00100155(struct FUN_0009ff84 *VAR_0, void *VAR_1) {
  int VAR_2;
  int VAR_3;
  void *VAR_4;
  VAR_2 = FUN_0009ff86(VAR_0, VAR_1);
  VAR_3 = 0;
  VAR_4 = FUN_0009ff88(VAR_0->VAR_5, VAR_2);
  while (VAR_4) {
    if (VAR_0->VAR_6 < VAR_3) { return -1; }
    if (!VAR_0->VAR_7(VAR_4, VAR_1)) { break; }
    VAR_2 = FUN_0009ff89(VAR_0, VAR_2);
    VAR_4 = FUN_0009ff88(VAR_0->VAR_5, VAR_2);
    VAR_3++;
  }
  if (VAR_4) { return VAR_2; }
  return -1;
}
  \end{lstlisting}
  \caption{Figure~\ref{fig:intro_original}, having been compiled, decompiled and run through LLM4Decompile-Ref~\cite{llm4decompile}.
  While the predicted code represents the same sequence of operations as \ref{fig:intro_original}, it is difficult to know if the prediction is correct without knowing the types of the fields used and how they're arranged within the type \cinline{struct FUN_0009ff84}.
  LLM4Decompile's solution is \emph{underspecified}; the semantics are not well-defined without the definition of \cinline{struct FUN_0009ff84}.\label{fig:intro_llm4decompile}}
  \end{subfigure}
  \hfill
  \begin{subfigure}[b]{0.49\textwidth}
  \begin{lstlisting}[style=cstyle,basicstyle=\ttfamily\bfseries\scriptsize]
struct hash_t {
  int size;
  int count;
  struct hash_table_t *table;
  int (*hash)(void *key);
  int (*cmp)(void *key1, void *key2);
};
struct hash_table_t {
  int size;
  void **items;
};
  
int hash_find(struct hash_t *hash, void *key) {
    int index = hash_index(hash, key);
    int i = 0;
    void *item = hash_get(hash->table, index);
    while (item != ((void *)0)) {
        if (i++ > hash->size) { return -1; }
        if (hash->cmp(item, key) == 0) { break; }
        index = hash_next(hash, index);
        item = hash_get(hash->table, index);
    }
    return (item == ((void *)0)) ? -1 : index;
}
  \end{lstlisting}
  \caption{Figure \ref{fig:intro_original}, having been compiled, decompiled, and run through \idioms.
  Unlike Figure~\ref{fig:intro_llm4decompile}, the \idioms model jointly predicts the necessary type definitions along with the function.\label{fig:intro_idioms}}
  \end{subfigure}
  
\caption{A function with two user-defined types and different decompilations of it.}
\end{figure*}

However, the neural decompiler output is still far from ideal.
A security practitioner glancing at the original code (Figure~\ref{fig:intro_original}) can immediately tell that the function is hashing-related because of its variable names (\cinline{hash_find_index}) or output type (\cinline{struct hash}).  The neural-decompiled code (Figure~\ref{fig:intro_llm4decompile}) contains no such obvious clues.
More problematically, 
reverse engineers often work across multiple levels of abstraction~\cite{votipka2020observational} such as assembly code and decompiled code.
This requires knowledge of the memory layout of the data structures in the decompiled code. 
Although \cinline{struct FUN_0009ff84} is clearly a structure, existing neural decompilers \emph{are not trained to produce type definitions.} Its memory layout is thus unknown.  
Type definitions are also required for compilation, and most types of static analysis, such as those that find and patch vulnerabilities, or deploy certain types of software defenses.
This limitation is significant.
As we discuss in Section~\ref{sec:dataset_generation} (Table~\ref{tab:dataset_complexity}), user-defined types (UDTs), such as structs, are widespread in real code.
This problem is largely masked because existing research benchmarks feature very few, if any, UDTs.

In this work, we propose a novel method for training neural decompilers that explicitly reconstructs full user-defined type definitions alongside reconstructed code.  
Figure~\ref{fig:intro_idioms} shows the output of our approach, which we call \idioms, since it recovers idiomatic code.
\idioms predicts a complete type definition for the output structure, \cinline{struct hash_t}, as well as the names of its fields.  The resulting code is well-defined and thus amenable to the types of reasoning common in reverse engineering. 
\idioms is based on two main insights:

\vspace{1ex}
\noindent\emph{Insight 1: Code and type definitions should be predicted jointly.}
Decompiled code and type definitions are fundamentally interdependent: meaningful variable names and field accesses depend on understanding the underlying data structures, while accurate type reconstruction requires knowing how those types are used throughout the code.
This interdependence argues strongly for joint prediction rather than sequential approaches.
Existing neural decompilers that predict code without type definitions produce \emph{underspecified} outputs---the semantics of struct field accesses cannot be determined without knowing the memory layout of those structs.
Conversely, predicting types in isolation from their usage context discards valuable information about how fields are accessed and manipulated.
An alternative approach might be to apply existing type recovery tools~\cite{dirty,tygr,resym} as a post-processing step to neural decompiler output.
However, this sequential approach has fundamental limitations.
First, existing type recovery tools are designed for executables or deterministic decompiler output, not the often-uncompilable code produced by neural decompilers.
Second, and more importantly, this approach prevents the neural decompiler from leveraging type information during code generation.
When field names and their usage are predicted jointly, the model can maintain consistency between how a field is defined (e.g., \cinline{hash_size}) and how it is used throughout the function (e.g., \cinline{h->hash_size}).
This type of consistency is only possible with joint prediction.

\vspace{0.5ex}
\noindent\emph{Insight 2: Scattered evidence for UDT type recovery necessitates consideration of broad context.} 
Predicting the definition of a UDT is fundamentally difficult because of what we call the \emph{scattered evidence} problem: typically, only a subset of a UDT's fields are accessed within any given function~\cite{tiesurvey}. This incomplete view makes type inference challenging when considering functions in isolation, as the full structure and semantics of UDTs cannot be determined from partial usage patterns.
This challenge mirrors the fundamental problem faced by traditional type inference algorithms when analyzing executables---they must aggregate evidence across multiple functions to reconstruct complete type definitions, which is precisely why such algorithms are inherently interprocedural~\cite{tiesurvey}.

To address this challenge, we provide broader context in the form of \emph{neighboring} functions---those close to the target function in the call graph. A function's callees and callers process related input, output, and internal values, often revealing additional clues about UDT structure and usage patterns. This interprocedural evidence enables more accurate type inference by aggregating partial information scattered across the program's call graph, allowing the model to reconstruct complete UDT definitions that would be impossible to infer from any single function alone.

Based on these insights, we
propose a new training strategy and associated family of \idioms neural
decompilation models that (a) jointly predict code and type definitions,
enabling consistent type application, and (b) leverage neighboring functions in
the call graph to provide the necessary context for UDT
reconstruction.  
Like LLM4Decompile-Ref~\cite{llm4decompile} and earlier work on specific renaming tasks~\cite{dire,dirty}, our
models take as input the output of deterministic decompilers applied to
target binaries. This approach leverages decades of progress in
deterministic decompilation and reduces the difficulty of the model's task.
Unlike existing neural decompilers~\cite{slade,llm4decompile,nova}, \idioms
models explicitly predict complete definitions of all UDTs used by a
function alongside the function definitions. Unlike some prior
work~\cite{slade}, \idioms does not require test cases, which are typically unavailable
in reverse engineering scenarios. 

We demonstrate experimentally that \idioms substantially outperforms existing work: \idioms scores
17--36\% better than LLM4Decompile~\cite{llm4decompile} and 37--78\% better than
Nova~\cite{nova} on \exebench~\cite{exebench}, the most challenging existing
benchmark, and 95--205\% better than either on more realistic code with real
UDTs, even using the most permissive evaluation metric.

We additionally perform controlled experiments to demonstrate \emph{why} \idioms performs well.  First, we show that neurally
decompiling functions containing UDTs is substantially more challenging than
decompiling those without UDTs, even when controlling for function size.  This finding generalizes across model sizes and
architectures.  We also demonstrate that including neighboring context substantially improves the structural accuracy of predicted UDTs by up to 63\%, with larger models benefiting more.

In summary, we contribute:
\begin{itemize}
\item A new dataset, \realtype, containing 154,301 training functions and 2,862 evaluation functions with realistic user-defined types (UDTs) and their complete definitions extracted from preprocessed source code.
\item A novel approach that enables neural decompilers to jointly predict both function code and complete user-defined type definitions simultaneously.
\item The \idioms family of neural decompilation models that achieve state-of-the-art performance, outperforming LLM4Decompile and Nova by 95-205\% on realistic code with UDTs.
\item Experimental evidence demonstrating that (a) UDTs substantially increase the difficulty of neural decompilation, (b) neighboring function context significantly improves UDT prediction accuracy by up to 63\%, and (c) these findings generalize across model architectures and sizes.
\end{itemize}

In the interest of open science, we release the code used to build the datasets, train and evaluate the models, as well as our dataset, \realtype, all \idioms models, and other models we train in our experiments for comparison.\footnote{Code can be found at \url{https://github.com/squaresLab/idioms}; models and datasets can be found at \url{https://doi.org/10.5281/zenodo.14797016}.}

\section{Approach}
\label{sec:methodology}

\begin{figure*}
\includegraphics[width=\textwidth]{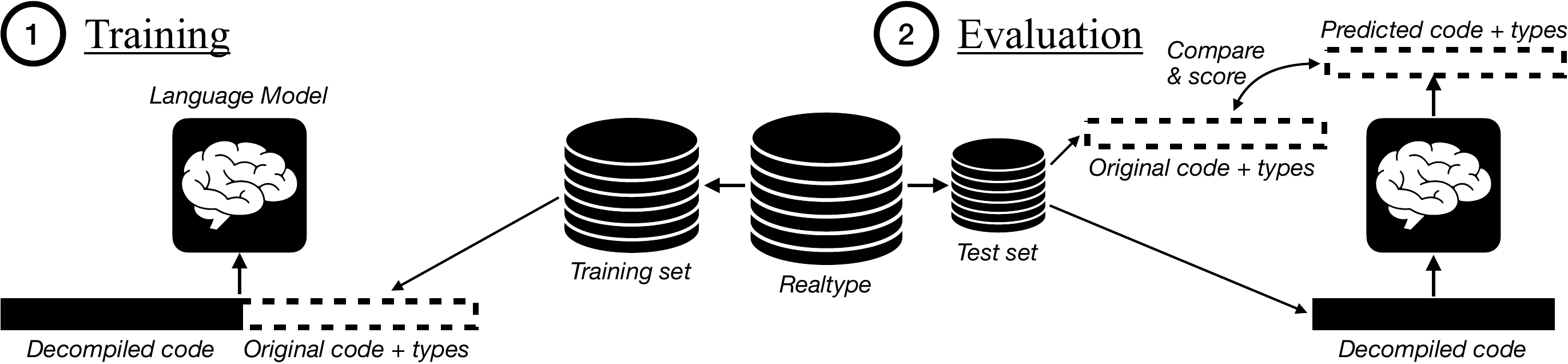}
\caption{Our high-level approach. We use our dataset, \realtype, to finetune causal language models into \idioms models then evaluate them.\label{fig:approach}}
\end{figure*}

Figure~\ref{fig:approach} overviews our approach.  
Decompilation aims to produce source or source-like representations of compiled
code, to help reverse engineers more easily understand or manipulate it.  
Neural decompilation generally entails \emph{training} machine learning models
to predict the original source code for some target binary function of
interest
(left-hand-side of Figure~\ref{fig:approach}).
\idioms models take as input the output of a deterministic decompiler, and
predicts the original
source code definition of a target function, along with a list of user-defined
types in that function.
\idioms operates on decompiler output rather than on the compiled binary directly because doing so naturally leverages significant advances in deterministic decompilation.
Tan et al.~\cite{llm4decompile} compare neural decompilation from assembly and from deterministic decompiler output and find that the latter produces correct output more often.
Section~\ref{sec:modeling} details the modeling approach, which entails finetuning pretrained causal language models, using call graphs to select context to inform predictions. 

Training models this way requires suitable example input and output data.
Existing datasets are inadequate for this task for two reasons: (1) a lack of
variables with UDTs and their definitions, and (2) they are function level, and
thus we cannot build call graphs. 
We therefore build a suitable new dataset, \realtype (Section~\ref{sec:dataset_generation}).

Once trained, models can predict the original code and UDT definitions
for a given decompiled function (right-hand-side of Figure~\ref{fig:approach}).
In practice, this can be done on any compiled code by applying a
deterministic decompiler first (standard, in reverse engineering) and then
applying an \idioms model to the result.  
For evaluation purposes, we  can take advantage of the ground truth available in
our data (the original code, pre-compilation). However, care must be taken to
evaluate on data not included in the training dataset. 

Our research questions
motivate training/finetuning multiple models, so we describe training details in that context. 

\subsection{Modeling}
\label{sec:modeling}

\begin{figure*}
\includegraphics[width=\textwidth]{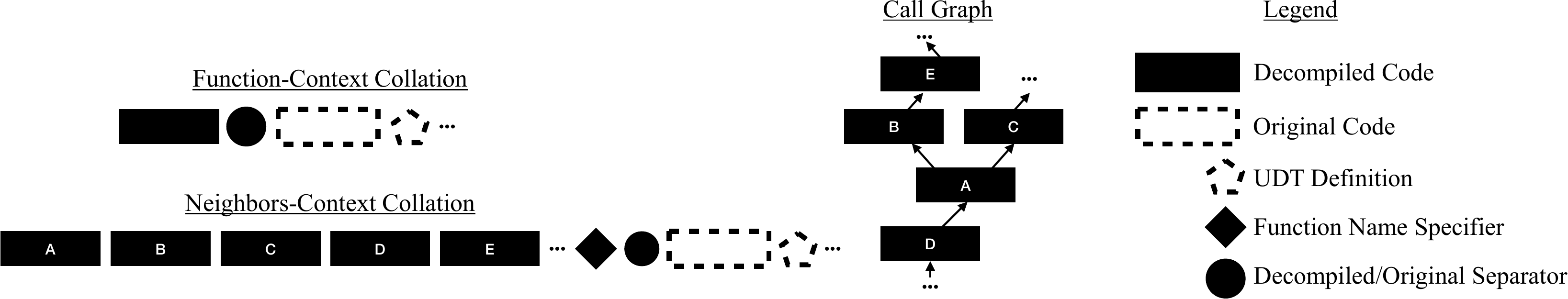}
\caption{Organization of the model training sequences. \idioms models use
neighboring context, which lists decompiled functions starting from a target
function in the context window, following a breadth-first search of the call
graph. In the figure, A is the target function. Function context is
used in some experiments to compare with
\idioms.\label{fig:context_organization}}
\end{figure*}

\noindent\textbf{Model architecture and finetuning.}
State-of-the-art language models are neural networks, machine learning models that encode information in real numbers called parameters that are learned via training. 
To partially overcome the challenges of developing sufficient training data for a particular task, a model can be \emph{pretrained} on data for a related task, and then \emph{finetuned} on a smaller quantity of data relevant to the target task.
We follow this paradigm for neural decompilation by finetuning causal language models pretrained on code.

Causal language models process token sequences; 
Figure~\ref{fig:context_organization} illustrates the organization of the
sequence we provide our model.
In this kind of supervised context, the task is to learn to predict an output
(original code + UDT definitions) from an input sequence (decompiled code). For
training, the input and output are concatenated into a single sequence,
delimited with a \emph{separator token}.
For evaluation, or use in practice, the model is prompted with the input and
separator token alone to generate the output (the rest of the sequence).

\vspace{0.5ex}
\noindent\textbf{Jointly predicting code and type definitions.}
The ``output'' side of the (training) context contains the function definition and
associated user defined types from the original source code. In Figure~\ref{fig:context_organization}, this is
everything to the right of \scalebox{1.5}{$\bullet$}.
Importantly, code and type definitions are in the same sequence.
Causal language models use the entire preceding sequence to predict the next token.
This means that field names, and how they are used in the predicted code, are
all 
available to the model as it generates type definitions. 
That is, code and type definitions are generated \emph{jointly}.

\vspace{0.5ex}
\noindent\textbf{Additional context}.
Evidence for UDT type recovery is scattered across multiple
functions~\cite{dramko2024taxonomy}, not just a given target for decompilation.
\idioms models therefore use neighboring function context in its prediction, to
partially address this problem. 
In Figure~\ref{fig:context_organization}, this context is to the left of the \scalebox{1.5}{$\bullet$}.
It is arranged such that the target decompiled code is first; callees and callers of the target function are placed subsequently.
Their callees and callers are then added (if not previously included), and so
on, up to the model context window size.
This breadth-first order is deliberate. A model should relate information
it learns from neighboring context to inform type prediction in the
target function. 
That information can only \emph{be} related if there's an unbroken trace from
the information source to the target function.
Additionally, there is a practical limit to the length of sequences models can handle. 
Providing call graph context in BFS order allows an effectively arbitrary cutoff on the right side
to respect context length constraints, while maintaining the integrity of the
trace from a given context function back to the target. 
This neighboring context is followed by a special name-indicator token, and then
tokens of the decompiled target function's name in the decompiled code.

Figure~\ref{fig:context_organization} also shows sequences with function-level context,
used by existing prior work on neural
decompilation~\cite{llm4decompile,nova,slade,katz2018,coda,cao2022boosting,hosseinibeyond}. However, sometimes due to architectural differences, decompiled and original
code are in separate sequences.
Function-level context consists of the target decompiled function.
We demonstrate the value of neighboring context (Section~\ref{sec:rq5}) by comparing with models built using only 
the decompiled function as context.

\subsection{Dataset}
\label{sec:dataset_generation}

Supervised machine learning requires training data. Our model architecture and overall goals impose two
requirements on a training dataset: it must contain code that uses representative user-defined
types, and we must be able to construct interprocedural call graphs from the
compiled binaries. 
Existing neural decompilation datasets like \exebench~\cite{exebench} and
\textsc{Humaneval-Decompile}~\cite{llm4decompile} provide only single-function context, and have fewer and
simpler UDTs than real-world code, if any (as we discuss below). 
We therefore construct a novel dataset, \realtype, to meet these requirements.

\vspace{0.5ex}
\noindent\textbf{Mining functions with UDTs.}
We cloned and compiled majority-C-language repositories from GitHub using the GitHub Cloner and Compiler (GHCC) tool~\cite{ghcc}.
The tool executes standard build configuration scripts, executes standard build scripts like Makefiles, and extracts any resulting ELF-format binary files, then archives the repository if it is under 100 MB in size.
Among the ELF files are object (\cinline{.o}) files.
Including object files increases the amount of data available because not all projects build completely.
However, it also means that for projects that do build completely, the same
function is present at least twice, at least in the object file and the complete
binary. 
We filter out all object files for which its corresponding functions appear in
another ELF-format binary in the same repository. 
For convenience, we adapt dataset preprocessing scripts from the DIRTY~\cite{dirty} replication package to interface
with Hex-Rays to decompile each binary.
We filter out PLT stubs, then canonicalize the function names to \cinline{funcX}, where X is an integer, a standard canonicalization scheme~\cite{hext5,nova,llm4decompile}.

To collect the original code and its associated types (our output data), we
modified GHCC to run \cinline{gcc}'s preprocessor. 
We parsed the preprocessed code to track \cinline{typedef} aliases and record
UDT definitions.
We store all types in a python object model representing the C type system built
on top of DIRTY's~\cite{dirty} type system, allowing fine-grained programmatic
type analysis.
The types we collect reflect the expressive power of C types, including
arbitrary nesting and typed function pointers. 

Next, we extract and textually canonicalize each function.
We traverse the function's AST to record all type descriptors (e.g. at variable declarations, typecasts, and the function's return type).
We de-alias each name by searching the previously created \cinline{typedef}
alias chains, replacing the names in text with a standard version. 
For instance, \cinline{int32} and \cinline{_int32_t} all become \cinline{int}.
\footnote{The C standard specifies the minimum size of each type, but not the
maximum, so technically \cinline{int} is ambiguous---hence the reason for the alternative type names in the first place. We performed all experiments
on the same platform, x86\_64 \texttt{linux} with \texttt{gcc}, so this was not an issue. We chose C
keywords for simplicity, but any alternative convention describing, e.g., type
sizes explicitly, would be fine, so long as it is consistent.}
Standard type forms makes results easier to evaluate and reduces ambiguity in
the mapping between a type's memory representation, and the syntactic form
learned by the neural decompiler. 
We store the canonicalized form of each function along with the type of each
variable within it.

We then matched preprocessed functions with decompiled functions and organize them by binary.
We do this so as not to leak information during training; a reverse engineer will attempt to analyze one or more binaries at once, without access to the original source.
Within each binary, we compute and store the call graph between functions.

\vspace{0.5ex}
\noindent\textbf{Deduplication.}
\emph{Data leakage} is a key risk in machine learning, because  
machine learning models tend to ``memorize'' examples in the training dataset.
Thus, it is important that training and evaluation sets be disjoint. 
The risk of data leakage via
pretraining is relatively low for neural decompilation (Section~\ref{sec:discussion}). However, the risk of data
leakage during finetuning remains, because
duplicates on GitHub are extremely common~\cite{spinellis2020dataset}.
Note that exact-match deduplication is insufficient: many copies of popular projects
represent different past versions of that project, or otherwise contain small
modifications.
We therefore deduplicate the dataset aggressively, so it can be conflidently
split into disjoint training and testing splits. 
Our approach is two-fold:
\begin{itemize}
\item Minhashing~\cite{minhashing}: Minhashing clusters similar text files, and is a popular choice to deduplicate code
datasets~\cite{TheStack,ThePile} due to its robustness.
We treat all C files in a repository as a ``document'' and treat resulting clusters of repositories as duplicates.
We include the repository from each cluster which produced the most data.
\item By-project splitting: we ensure that all data from a given repository is
assigned entirely to one of the train, validation, or test splits. 
This prevents models from memorizing project-specific details or conventions
(including UDTs) from some functions in a project, and then applying them to
other different functions in the same project.  
This has been demonstrated as empirically important in prior evaluations of ML-based decompilation~\cite{hext5}.
\end{itemize}
Each strategy targets a different type of data leakage, motivating their
combined use. 

\begin{table}
    \caption{Complexity in Evaluation Datasets\label{tab:dataset_complexity}}
    \centering
    \begin{tabular}{lrrr}
    \toprule
    {} & HEDecomp* & exebench & \realtype \\
    \midrule
    Lines of code & 15.2 & 13.9 & 14.2 \\
    Variables with a UDT (\%) & 0 & 0.5 & 26.4 \\
    Functions with a UDT (\%) &  0 & 1.9 & 53.4 \\
    Recursive UDT field count & N/A & 2.2 & 17.6\\
    Type-tree complexity & 1.4 & 1.5 & 16.2 \\
    UDT Type-tree complexity & N/A & 4.2 & 57.3 \\
    \bottomrule
    \end{tabular}
    \caption*{UDTs are user-defined types (\cinline{struct}, \cinline{union}).
    \realtype is our dataset. 
    \textsc{humaneval-decompile}~\cite{llm4decompile} (\textsc{HEDecomp*}) is based on programming challenge problems.
    \exebench~\cite{exebench} is also mined from GitHub, but contains many fewer UDTs. The type complexity metric measures the number of nodes in a tree representation of the type: primitive types are leaf nodes, all other modifiers are internal nodes.%
    UDT Type-tree complexity includes only UDT types, Type-tree complexity includes all types.
    The \exebench statistics are for its \cinline{test_real} partition. %
    }
    \end{table}

\vspace{0.5ex}
\noindent\textbf{\realtype characteristics.}
After deduplication, the \realtype dataset contained 154,301 training functions and 2,862 evaluation functions. \realtype has substantially more user-defined types than existing benchmarks.
Table~\ref{tab:dataset_complexity} compares \realtype with two popular benchmarks used to evaluate the state-of-the-art models LLM4Decompile~\cite{llm4decompile} and Nova~\cite{nova}.
They contain very few, if any, user-defined types, and those that are present are unrealistically simple.

\section{Experimental Design}
\label{sec:experimental_design}

Our experiments address five research questions:

\textbf{RQ1}: \emph{How does \idioms perform relative to the best existing techniques?}
How do the insights that enable \idioms give it an advantage over prior work?

\textbf{RQ2}: \emph{How does \idioms fare when exposed to optimizations?}
Optimizations are common in real code. Do \idioms' advantages hold up when dealing with optimized code?

\textbf{RQ3}: \emph{To what extent, if any, do UDTs increase the complexity of neural decompilation?}
Existing work is evaluated on benchmarks that contain limited UDTs.
But as we have discussed previously, predicting UDTs is valuable, and UDTs are commonplace in real code.
How much do UDTs impose an additional challenge for neural decompilers?

\textbf{RQ4}: \emph{How does \idioms performance change with model and dataset
size?}
We examine the overall performance of \idioms and how scaling up the process impacts that performance.

\textbf{RQ5}: \emph{How does adding neighboring context affect the quality of neural decompilers?}
The scattered evidence problem tells us that the evidence needed to understand UDTs may be spread throughout different functions of the program.
Does providing neighboring context help the model predict UDTs more effectively,
and if so, by how much?

\vspace{0.5ex}
\noindent\textbf{Baselines.} We compare against two state-of-the-art models as baselines: Nova~\cite{nova} and LLM4Decompile~\cite{llm4decompile}.
Nova is a neural decompiler that uses assembly as input rather than decompiled
C, and features a custom attention implementation designed for assembly.
There are two versions of LLM4Decompile, one which uses assembly as input and
one for code decompiled by Ghidra (an alternative to Hex-Rays). We use the
latter version, because it has the best performance.

We do not compare against one other piece of notable recent work in neural decompilation: SLaDe~\cite{slade}.
SLaDe achieves very strong results for a very small model size.
However, by design, SLaDe uses the test cases bundled with \exebench for each function during prediction.
Unfortunately, this is unrealistic in practice, because the source code was used
to create test cases, directly; even relaxing this assumption, reverse
engineering contexts often lack access to test suites, let alone high-coverage suites.
Even with source code, the authors of \exebench~\cite{exebench} could only
generate tests for 15.4\% of the dataset.

\subsection{Datasets and Processing}
\label{sec:datasets_and_baselines}

We use two datasets in our experiments: \realtype
(our new dataset), and \exebench, as a baseline.
\exebench is the most challenging existing benchmark used to evaluate existing
work; it also includes a number of example with generated test cases, which can
be used as a proxy for decompilation correctness.
It features code extracted from GitHub, along with definitions for external symbols associated with some functions.
Most of the definitions are synthetic and automatically generated by the authors.
They do not feature the full definitions of UDTs, however.
Many have a single field named \cinline{dummy}, marking it as a placeholder value.
A subset of the functions, about 15.4\% of the overall dataset, are associated
with automatically-generated unit test
cases. 
The test and validation sets are selected out of those 15.4\% of the functions for which tests could be generated (hence the bias illustrated in Table~\ref{tab:dataset_complexity}).

We used the compilable partition of \exebench for which deterministic
decompilation succeeded.
We evaluate \exebench experiments only on the ``real'' subpartition of
\exebench's test set, which avoids the above-noted problems; the ``synth''
subpartition is also easier, inflating performance~\cite{slade}.
We also exclude examples where the provided oracle solution and dependencies
don't compile, or don't pass all tests (about 16\% of the \texttt{test\_real} partition). 

We re-decompile both tests sets using Ghidra to evaluate LLM4Decompile; 
evaluating it directly on Hex-Rays data would represent a covariate shift that
would unfairly hurt LLM4Decompile's performance.

\vspace{0.5ex}
\noindent\textbf{Test set processing.}
We exclude the 16\% of the \exebench \texttt{test\_real} set on which our
decompilation scripts fail (resulting in no input for the model). 
Following LLM4Decompile~\cite{llm4decompile}, we do not evaluate on examples for which the decompiled function exceeds the model's context window.
This filters out about 3\% of each test set.
When comparing the impact of expanded context on prediction performance to the
prior work's use of function-only context, we evaluate both on the same subset
of the test set. 

In our experiments, all models are trained to predict the decompiled code first, followed by any user-defined types used in the function or in other user-defined types.
Additionally, model predictions on \realtype sometimes include degenerate text~\cite{holtzman2019curious}\footnote{This is a common but poorly-understood phenomenon, even in large and powerful models.} in the form of repeating type definition patterns, after predicting sensible function definitions and UDTs. We hypothesize this happens simply because the task---predicting complete UDT definitions, especially at the complexity found in \realtype---is extremely difficult. 
Because of the scattered evidence problem, it is sometimes case that the decompiled context is \emph{not} predictive of the full relevant UDT definition(s), so the model may just learn to generate arbitrary type definitions after anything it can figure out.
Degenerate text empirically always appears after correct predictions.
Fortunately, dealing with it is easy: in our evaluation, degenerate types unused by any variable are simply ignored, which can be done in practice as well.

\subsection{Metrics}
\label{sec:metrics}

A good neural decompiler (1) preserves execution semantics, and (2) 
restructures names, types, and code to be more readable and idiomatic, or written in a style that mimics a human developer's.
With respect to the state-of-the-practice, the former goal means that a good neural decompiler should be no worse than a deterministic one, and the latter means it must add value.
We use a suite of metrics to measure a neural decompiler's efficacy on both
fronts, comparing to the original source code as the gold standard; in some
respects, an ideal decompiler one that perfectly ``undoes'' the compilation.

\begin{figure}
\begin{subfigure}{0.49\columnwidth}
\begin{lstlisting}[style=cstyle,basicstyle=\ttfamily\bfseries\scriptsize]
int env_loc(char *s){
  int h = 0;
  while (*s++)
    h = h * 31 + *s;
  return h %
}
\end{lstlisting}
\caption{Original code}
\end{subfigure}
\begin{subfigure}{0.49\columnwidth}
\begin{lstlisting}[style=cstyle,basicstyle=\ttfamily\bfseries\scriptsize]
int hash(char *key) {
  int h = 0;
  while (*key++)
    h = h * 31 + *key;
  return h %
}
\end{lstlisting}
\caption{\idioms' prediction}
\end{subfigure}

\begin{subfigure}{\columnwidth}
\begin{lstlisting}[style=cstyle,basicstyle=\ttfamily\bfseries\scriptsize]
unsigned func0(const char *str) {
    unsigned hash = 0;
    while (*str)
        hash = hash * 31 + *str++;
    return hash %
}
\end{lstlisting}
\caption{Nova's prediction}
\end{subfigure}

\caption{A function which computes a variant of a polynomial rolling hash function, and its neural decompilation by \idioms and Nova~\cite{nova}. The original and \idioms' prediction have isomorphic dependency graphs, but Nova's does not.
The original code starts hashing the \emph{second} character in the string, while Nova's prediction starts with the first.
This is reflected in the function's dataflow dependencies.
  \label{fig:alignment_example}}

\end{figure}

\subsubsection{Semantic preservation}
Because program equivalence is undecidable, we seek a practical approximation of semantic preservation.  
We use complementary proxies: 

\vspace{0.5ex}
\noindent\textbf{(1) Unit test accuracy.}  \exebench includes unit tests with each example
  in the test set, which we leverage to evaluate the semantic fidelity of decompiled code.  
Tests are complete but unsound, but still a useful (if optimistic) proxy for correctness.
\emph{(Metric name: passes exebench tests)}

\vspace{0.5ex}
\noindent\textbf{(2) Static, dependency-based equivalence.}
\realtype does not include test cases, which are moreover complete but unsound.
Thus, we also use a static, dependency-based equivalence check to compare decompiled code to the original source~\cite{codealign}.  At a high-level, this approach asks ``Do two pieces of code perform the same operations, in the same order?'' It does so by comparing function dependency graphs, with nodes representing operations (assignments, function calls, etc.), data sources (constants, parameters, global variables), and edges control- and data-flow dependencies between them.  
``Inputs'' to the operations, like the functions' parameters and constants, are represented with their own nodes;
UDT fields are represented as constants.
If two such dependency graphs are isomorphic, this means the functions they represent have the same structure, performing the same operations in the same (partial) order, even if variable names differ.

Figure~\ref{fig:alignment_example} shows an example of dependency- equivalent and -nonequivalent functions, by this standard.  Figure~\ref{fig:alignment_example} shows a particularly subtle example of a common way that neural decompilers can be wrong; many of these manifest as incorrect dependencies.  Other incorrect predictions include obvious omissions, additions, reordered expressions or other incorrect representations of  function semantics.  dependency-equivalence is a good way to capture these mistakes.\footnote{Conversely, dependency-equivalence can fail to recognize different semantically-equivalent syntactic forms, like \cinline{x * 2} with \cinline{x << 1}. This is a common optimization, though neural decompilers are trained to predict the original code so they learn to \emph{undo} optimizations.}

Dependency graph isomorphism provides a sound approximation of program equivalence~\cite{yang1989detecting}, with two caveats:  side effects, a rare source of unsoundness in practice~\cite{codealign}; and type correctness.  We therefore also measure the fraction of a neural decompiler's predictions for which the prediction is dependency-equivalent to the corresponding ground truth function, \emph{and} where all variable types are equivalent to those in the ground truth.
Note that this type requirement is very conservative; vanilla dependency-equivalence captures whether the operations in a decompiled function are at minimum correct, and in the correct order. 

\emph{(Metric names: dependency-equivalence, strict dependency-equivalentce (typechecks).)}

Dependency-equivalence relies on UDT definitions to align field accesses,
and thus can be overly harsh on prior work, like
LLM4Decompile and Nova, that do not produce full type definitions.
However, these neural decompilers still often usefully predict variables to have
\cinline{struct} types, and rewrite pointer arithmetic into struct field-access
operations (as in Figure~\ref{fig:intro_llm4decompile}).
However, the \emph{names} of the fields often differ, making it difficult to determine which fields correspond to each other.
Thus, we introduce a more permissive version of dependency-equivalence under a principle of \emph{consistency}: there must be a bijective mapping between the names of the fields that occur in dependency-graph nodes that map together.
For instance the fragment \cinline{pt->x + pt->y} is consistent with \cinline{pt->a + pt->b} because there exists a bijective mapping between the field names: \cinline{x} $\leftrightarrow$ \cinline{a}, \cinline{y} $\leftrightarrow$ \cinline{b}.
However, no such mapping exists between \cinline{pt->x + pt->y} and \cinline{pt->a + pt->a}.
We do the same for function names.\footnote{There is no need for consistent (local) variable names since these manifest in dataflow dependencies.}
\emph{(Metric name: relaxed dependency-equivalent (consistency))}

\subsubsection{Code improvements}
It is also important to quantify how the neural decompiler has improved the code by adding abstractions like accurate variable names and types.

\vspace{0.5ex}
\noindent\textbf{Variable Names and Types.}
A challenge in evaluating variable name accuracy is determining which variables in the prediction correspond to those in the original source.
The prediction may break up expressions into smaller subexpressions, with results stored in intermediate variables, or vice versa.
This means that, except for function parameters, it can be ambiguous how to programmatically map between variables in the original and the decompiled versions of a function to compare their names and types.
To address this challenge, we identify the operations that map to one another in the previously-computed isomorphic maps computed for dependency-equivalence.
With this mapping, we compare variables' names and types.
\emph{(Metric names: variable name accuracy, variable type accuracy)}

\vspace{0.5ex}
\noindent\textbf{UDT accuracy.}
One of the key challenges with decompiling real C code is reconstructing user-defined types.
One novelty of the \idioms approach is that it predicts UDTs alongside the code, and thus we measure accuracy on UDTs separately. 
UDTs are often named, as are their fields; perfectly predicting both is challenging. 
That said, knowing and predicting a type's structure---the type of its fields, and their order---is still helpful to a reverse engineer, even if the names are imperfectly predicted. 
We therefore decompose UDT prediction accuracy, reporting the fraction of UDTs for which the structural layout matches (ignoring type and field names).
When computing our metric ``strict dependency-equivalentce (typechecks)'', we use structural equivalence in determining type equivalence.
\emph{(Metric names: UDT variable nominal accuracy, UDT variable structural accuracy)}
\vspace{8pt}

In summary, unit test accuracy is an upper bound on the rate at which neural decompilers preserve semantics, while strict dependency-equivalence forms an approximate lower bound.
Variable name accuracy, the three type accuracy measures (variable type accuracy, UDT variable nominal accuracy, and UDT variable structural accuracy), quantify the improvements that neural decompilers make to the code.
Restructuring code to be like the original is also an improvement, so dependency-equivalence can be thought of as an improvement metric as well.

\subsection{Setup}
\label{sec:methodlogy}

\noindent\textbf{Overall performance.} The first research question compares \idioms overall performance (trained on
the \realtype training set) to Nova~\cite{nova} and
LLM4Decompile~\cite{llm4decompile} on both \exebench and \realtype.
We use the \idioms models trained by fine-tuning the 7-billion parameter version of CodeGemma~\cite{codegemma} with a QLoRA~\cite{qlora} adapter, and compare to the 6.7b-sized versions of the related work.

\vspace{0.5ex}
\noindent\textbf{Compiler optimizations.}To evaluate how performance varies
with GCC optimization levels (RQ2), we train and
evaluate an \idioms model as well as similarly-sized versions of Nova and
LLM4Decompile when \realtype code is compiled by gcc at levels O0--O3.  
We disable inlining optimizations because it makes the predictions difficult to evaluate: our oracle, the original code, is not inlined, yet a neural decompiler will likely propagate the body of the inlined function to the prediction level.\footnote{Prior work evaluates on optimizations~\cite{slade,llm4decompile,nova} but these works use simpler datasets which consider each function in isolation, so inlining is not possible.}

Our dataset creation script did not always succeed at all optimization levels for each decompiler.  
For fair comparison across optimization levels, we include only functions that successfully decompiled at all levels (O0-O3).  This reduces the test set by 21\% for HexRays, but 76\% for Ghidra.  Because the resulting Ghidra set is problematically small, for the Ghidra-based evaluation of LLM4Decompile, we use  the subset of Hex-Rays-decompilable functions that also decompiled in Ghidra for the test set. 
This means that the Ghidra test sets for the compiler optimization results are not uniform.  However, as these are test sets, not training, the impact on results is negligible. 

\vspace{0.5ex}
\noindent\textbf{Model size.} To evaluate the degree to which our innovations are applicable to a variety of
model types and sizes, we finetune \idioms models from  
five pretrained models:
\begin{itemize}
\item CodeQwen2.5~\cite{qwen}, 0.5 billion parameter version
\item LLM4Decompile~\cite{llm4decompile}, 1.3 billion parameter version
\item CodeGemma~\cite{codegemma}, 2 billion parameter version
\item CodeGemma~\cite{codegemma}, 7 billion parameter version
\item CodeLlama~\cite{codellama}, 7 billion parameter version.
\end{itemize}
We use the common convention in machine learning to attach \cinline{-Xb} to the
name of each model, where \cinline{X} is the number of parameters in that model,
in billions.

We perform traditional finetuning on the smallest model, CodeQwen2.5-0.5b.
Finetuning becomes computationally prohibitive 
for models above 1 billion parameters in size; we therefore leverage recent results using
adapters and quantization (QLoRA~\cite{qlora} adapters), allowing for a
high-fidelity, computationally tractable approximation of full finetuning.
We provide additional training details, for replicability, in the Appendix, and
our replication package. 

\vspace{0.5ex}
\noindent\textbf{Ablation.}
Overall \idioms performance is evaluated by training models on \realtype with
all features.  
However, both the nature and size of training datasets can impact model
performance.  
We evaluate the effect of both, as well as the effect of our design choices, by training and evaluating alternative models
that allow for controlled comparisons:
\begin{itemize}
\item \textbf{exebench}: This experimental
setting trains and evaluates a neural decompilation model on \exebench, the most
complex benchmark in prior work. 
Because \exebench does not support interprocedural callgraphs, this
version of \idioms still jointly predicts names and UDT definitions, but does so without neighboring context. 
\item \textbf{parity-exebench}:
\exebench (2,383,839 functions) is substantially larger than \realtype (154,301
functions). To control for training set size in model performance, we subsampled
\exebench to create a smaller training set that matched \realtype in size for
training.  Comparing these results to the \textbf{exebench} setting controls for
training dataset size.
\item \textbf{functions-realtype}: To assess the effect neighboring function context
on \idioms performance, we train a version of \idioms with only
function context on the \realtype dataset.  Comparing these results to the full
\idioms results controls for the effect of the neighboring context design
decision. Additionally, recall that models trained on
\exebench by necessity only provide function context. Thus, this setting varies
from \textbf{parity-exebench} by dataset \emph{composition} only. We thus compare
these two settings to reveal the effect of training on \exebench versus the
more-realistic \realtype.  
\end{itemize}

We perform these ablations across all fine-tuned models, as described above.

\section{Results}
\label{sec:results}

\begin{table*}[t]
\begin{subtable}{0.5\textwidth}
\caption{Relaxed dependency-equivalence (consistency)\label{tab:consistently_aligned}}
\begin{tabular}{lr|rrrr}
\toprule
& & \multicolumn{4}{c}{\realtype} \\
&exebench&O0&O1&O2&O3\\
\midrule
\idioms& \textbf{34.1} & \textbf{32.3} & \textbf{28.0} & \textbf{26.2} & \textbf{25.5}\\
LLM4Decompile & 27.9 & 10.6 & 6.2 & 6.3 & 5.9\\
Nova & 24.8 & 16.6 & 9.4 & 8.0 & 7.5\\
\bottomrule
\end{tabular}
\end{subtable}
\hfill
\begin{subtable}{0.5\textwidth}
\caption{Dependency-equivalence\label{tab:perfectly_aligned}}
\begin{tabular}{lr|rrrr}
\toprule
& & \multicolumn{4}{c}{\realtype} \\
&exebench&O0&O1&O2&O3\\
\midrule
\idioms & \textbf{33.7} & \textbf{21.6} & \textbf{19.1} & \textbf{18.0} & \textbf{18.0}\\
LLM4Decompile & 27.4 & 7.3 & 3.6 & 4.6 & 5.1\\
Nova & 23.9 & 6.7 & 5.6 & 5.0 & 4.7\\
\bottomrule
\end{tabular}
\end{subtable}

\vspace{4pt}

\begin{subtable}{0.5\textwidth}
\caption{Strict dependency-equivalence (typechecks)\label{tab:perfectly_aligned_and_typechecks}}
\begin{tabular}{lr|rrrr}
\toprule
&exebench&O0&O1&O2&O3\\
\midrule
\idioms & \textbf{23.9} & \textbf{9.8} & \textbf{8.6} & \textbf{7.2} & \textbf{7.0}\\
LLM4Decompile & 17.6 & 3.3 & 2.6 & 3.5 & 3.7\\
Nova & 13.4 & 0.9 & 2.5 & 2.7 & 2.5\\
\bottomrule
\end{tabular}
\end{subtable}
\begin{subtable}{0.25\textwidth}
\caption{passes \exebench tests\label{tab:passes_exebench_tests}}
\begin{tabular}{lr}
\toprule
&exebench\\
\midrule
\idioms & \textbf{54.4}\\
LLM4Decompile & 46.3\\
Nova & 37.5\\
\bottomrule
\end{tabular}
\end{subtable}

\vspace{4pt}

\begin{subtable}{0.5\textwidth}
\caption{Variable name accuracy\label{tab:variable_name_accuracy}}
\begin{tabular}{lr|rrrr}
\toprule
&exebench &O0&O1&O2&O3\\
\midrule
\idioms & \textbf{20.6} & \textbf{19.8} & \textbf{18.7} & \textbf{17.9} & \textbf{17.8}\\
LLM4Decompile & 14.7 & 3.4 & 3.2 & 3.9 & 3.5\\
Nova & 12.9 & 4.5 & 3.2 & 3.5 & 3.5\\
\bottomrule
\end{tabular}
\end{subtable}
\hfill
\begin{subtable}{0.5\textwidth}
\caption{Variable type accuracy\label{tab:variable_type_accuracy}}
\begin{tabular}{lr|rrrr}
\toprule
&exebench&O0&O1&O2&O3\\
\midrule
\idioms & \textbf{58.2} & \textbf{38.3} & \textbf{34.0} & \textbf{33.3} & \textbf{32.6}\\
LLM4Decompile & 45.5 & 13.4 & 11.4 & 11.7 & 10.7\\
Nova & 41.8 & 13.0 & 9.9 & 11.5 & 12.1\\
\bottomrule
\end{tabular}
\end{subtable}

\vspace{4pt}

\begin{subtable}{0.5\textwidth}
\caption{UDT variable nominal accuracy\label{tab:variable_udt_exact_matches}}
\begin{tabular}{lr|rrrr}
\toprule
&exebench&O0&O1&O2&O3\\
\midrule
\idioms & \textbf{20.7} & \textbf{6.4} & \textbf{5.6} & \textbf{6.0} & \textbf{5.7}\\
LLM4Decompile & 0.0 & 0.0 & 0.0 & 0.0 & 0.0\\
Nova & 0.0 & 0.0 & 0.0 & 0.0 & 0.0\\
\bottomrule
\end{tabular}
\end{subtable}
\hfill
\begin{subtable}{0.5\textwidth}
\caption{UDT variable structural accuracy\label{tab:variable_udt_composition_matches}}
\begin{tabular}{lr|rrrr}
\toprule
&exebench&O0&O1&O2&O3\\
\midrule
\idioms & \textbf{34.5} & \textbf{15.0} & \textbf{14.7} & \textbf{13.4} & \textbf{12.3}\\
LLM4Decompile & 0.0 & 0.0 & 0.0 & 0.0 & 0.0\\
Nova & 0.0 & 0.0 & 0.0 & 0.0 & 0.0\\
\bottomrule
\end{tabular}
\end{subtable}
\caption{Performance of \idioms, Nova~\cite{nova} and
LLM4Decompile~\cite{llm4decompile}) on \exebench~\cite{exebench} and a subset of
\realtype that decompiled at all
optimization levels. All values are percentages; higher is better. 
Nova and LLM4Decompile score 0 on UDT-related metrics
because they do not predict UDTs. (UDT metric scores are computed on the set of
variables that have UDTs in the original code.)\label{fig:work_comparison}}
\end{table*}

Table~\ref{fig:work_comparison} shows results for RQ1, comparing \idioms
performance to prior work, and RQ2, evaluating the impact of compiler
optimization levels.  We discuss both questions in
Section~\ref{sec:rq1-results}. 

Results to support the remaining RQs are show in Table~\ref{fig:main_results},
which is organized by model and by training configuration.  The table is
structured such that each column differs by one key experimental setting from
adjacent settings, allowing for controlled comparisons.  
The \textbf{exebench} and \textbf{parity-exebench} columns differ by training size (the first is large); \textbf{parity-exebench} and \textbf{functions-realtype} differ by dataset type (\exebench vs. \realtype) and thus the complexity and number of UDTs in both training and evaluation; \textbf{functions-realtype} and \idioms differ by the context provided to the model (decompiled function only vs. decompiled function and neighboring functions). 

Section~\ref{sec:rq3-results}
discusses the 
impact that realistic UDTs have on the complexity on the neural decompilation
task, both in training and evaluation; Section~\ref{sec:rq4-results} addresses
model size and scaling; and Section~\ref{sec:rq5} evaluates the use of
neighboring function context in \idioms design. 

Note that, in general, there is high variance in UDT metrics for the \exebench-based
experiments because there very few UDTs in the \texttt{test\_real} subset of
\exebench.

\subsection{Decompilation performance (RQ1 and RQ2)}
\label{sec:rq1-results}

Table~\ref{fig:work_comparison} shows performance of \idioms, LLM4Decompile, and
Nova on \exebench and \realtype, including performance at different compiler
optimization levels. The results are organized by performance metrics. 

\vspace{0.5ex}
\noindent\textbf{Overall performance.} \idioms substantially outperforms existing work.
On \exebench, \idioms scores 17-36\% higher than LLM4Decompile and 38-78\%
better than Nova on all correctness metrics, while scoring similarly highly on
code improvements.
This improvement is despite the fact that the \exebench experiments do not
leverage \idioms' key differentiators.
One possible reason is that Nova and LLM4Decompile both bear the hallmarks of being trained on code where the function names are left in the decompiled code (via debug information).
In particular, both tend to copy the generic function name (e.g. \cinline{FUN_00100155} or canonicalized into \cinline{func0}) from the input to the output unchanged---a pattern learned when the input (decompiled code or assembly) and output (original code) have the same name in training.
Indeed, sample training data linked to from the LLM4Decompile repository has function names in the decompiled code.
Function names confer a large amount of information about the function to neural
models~\cite{jin2023binary}, rendering the overall task substantially easier. 
Except where dynamic linking is required, function names in binaries are uncommon in practice, so we evaluate on code with the names removed.
Other confounding factors include model architecture and input type---anecdotally, we find Hex-Rays' output to be better than Ghidra's, for instance.

However, the performance gaps are much larger on \realtype, where the innovations that differentiate \idioms from prior work are most relevant.
Notably, Nova and LLM4Decompile score 0 on the UDT-related metrics (Tables~\ref{tab:variable_udt_composition_matches} and \ref{tab:variable_udt_exact_matches}) because they do not predict any user-defined type definitions at all.
But UDTs and the code that interacts with them are intrinsically linked.
\idioms scores 95-205\% higher than Nova and LLM4Decompile on correctness metrics---even the most permissive that does not require UDT definitions (Table~\ref{tab:consistently_aligned}).
In turn, variable name and type metric scores are low partially because there are many variables in the original code that don't correspond to anything in the nonequivalent predicted code.

\vspace{0.5ex}
\noindent\textbf{Impact of compiler optimizations.}
Table~\ref{fig:work_comparison} also delineates results by gcc optimization
level; these reults entail training and testing on \realtype. 
In line with related work~\cite{nova,llm4decompile,slade} on easier datasets, \idioms' performance degrades somewhat as optimization levels increase.
There is a large drop at O1, and smaller drops when other optimizations are applied.
Still, \idioms performs much better at O3 than related work does at
O0: 25.5\% on relaxed dependency equivalence vs. 16.6\%.

\begin{tcolorbox}[width=\columnwidth,
                  boxsep=0pt,
                  left=4pt,
                  right=7pt,
                  top=7pt,
                  arc=4pt,
                  boxrule=1pt,
                  toprule=1pt,
                  colback=white
                  ]%
\textbf{Takeaway}:  \idioms greatly outperforms state-of-the-art neural decompilers, especially on realsitic code with UDTs.  Performance drops sharply at O1, but higher levels of optimization have little additional effect.
\end{tcolorbox}

\begin{table*}
\begin{subtable}{1.0\textwidth}
\centering
\caption{CodeQwen2.5-0.5b\label{tab:qwen_0.5b}}
\begin{tabular}{clrr|rr}
\toprule
&&exebench&parity-exebench&functions-realtype& \idioms\\
\midrule
\parbox[t]{0mm}{\multirow{4}{*}{\rotatebox[origin=c]{90}{\scriptsize Correctness}}}
&Relaxed dependency-equivalence (consistency)& 30.8 & 26.0 & 24.2 & 22.3\\
&Dependency-equivalence & 30.6 & 26.0 & 16.2 & 15.6\\
&Strict dependency-equivalence (typechecks) & 20.4 & 17.0 & 7.1 & 7.2\\
&Passes \exebench tests & 44.4 & 32.1 & -- & --\\
\hline
\parbox[t]{0mm}{\multirow{4}{*}{\rotatebox[origin=c]{90}{\scriptsize Improvement}}}
&Variable name accuracy & 18.8 & 15.2 & 13.8 & 13.6\\
&Variable type accuracy & 54.4 & 45.6 & 32.9 & 30.4\\
&UDT variable nominal accuracy& 11.1 & 6.9 & 3.2 & 3.7\\
&UDT variable structural accuracy & 37.0 & 10.3 & 6.0 & 6.4\\
\bottomrule
\end{tabular}
\end{subtable}
\vspace{4pt}

\begin{subtable}{1.0\textwidth}
\centering
\caption{LLM4Decompile-1.3b-v2\label{tab:llm4decompile_1.3b}}
\begin{tabular}{clrr|rr}
\toprule
&&exebench&parity-exebench&functions-realtype&  \idioms \\
\midrule
\parbox[t]{0mm}{\multirow{4}{*}{\rotatebox[origin=c]{90}{\scriptsize Correctness}}}
&Relaxed dependency-equivalence (consistency)& 32.1 & 29.3 & 26.9 & 26.8\\
&Dependency-equivalence & 31.9 & 29.1 & 18.0 & 17.3\\
&Strict dependency-equivalence (typechecks) & 23.3 & 19.8 & 8.4 & 8.3\\
&Passes \exebench tests &49.1 & 41.3 & -- & --\\
\hline
\parbox[t]{0mm}{\multirow{4}{*}{\rotatebox[origin=c]{90}{\scriptsize Improvement}}}
&Variable name accuracy & 19.2 & 17.2 & 16.1 & 16.9\\
&Variable type accuracy & 54.9 & 49.6 & 36.1 & 35.3\\
&UDT variable nominal accuracy& 13.8 & 12.0 & 3.8 & 4.6\\
&UDT variable structural accuracy & 41.4 & 28.0 & 10.0 & 11.1\\
\bottomrule
\end{tabular}
\end{subtable}
\vspace{4pt}

\begin{subtable}{1.0\textwidth}
\centering
\caption{CodeGemma-2b\label{tab:codegemma_2b}}
\begin{tabular}{clrr|rr}
\toprule
&&exebench&parity-exebench&functions-realtype& \idioms \\
\midrule
\parbox[t]{0mm}{\multirow{4}{*}{\rotatebox[origin=c]{90}{\scriptsize Correctness}}}
&Relaxed dependency-equivalence (consistency)& 31.1 & 29.2 & 26.5 & 26.1\\
&Dependency-equivalence & 30.9 & 29.0 & 17.0 & 18.1\\
&Strict dependency-equivalence (typechecks) & 21.5 & 21.1 & 8.5 & 8.6\\
&Passes \exebench tests & 48.3 & 41.3 & -- & --\\
\hline
\parbox[t]{0mm}{\multirow{4}{*}{\rotatebox[origin=c]{90}{\scriptsize Improvement}}}
&Variable name accuracy & 19.5 & 16.9 & 15.7 & 17.6\\
&Variable type accuracy & 55.6 & 50.6 & 36.1 & 35.4\\
&UDT variable nominal accuracy & 24.1 & 24.1 & 3.8 & 4.9\\
&UDT variable structural accuracy &55.2 & 41.4 & 7.4 & 11.5\\
\bottomrule
\end{tabular}
\end{subtable}
\vspace{4pt}

\begin{subtable}{1.0\textwidth}
\centering
\caption{CodeGemma-7b\label{tab:codegemma_7b}}
\begin{tabular}{clrr|rr}
\toprule
&&exebench&parity-exebench&functions-realtype& \idioms \\
\midrule
\parbox[t]{0mm}{\multirow{4}{*}{\rotatebox[origin=c]{90}{\scriptsize Correctness}}}
&Relaxed dependency-equivalence (consistency)& 34.1 & 31.7 & 28.5 & 27.5\\
&Dependency-equivalence & 33.7 & 31.3 & 18.1 & 18.3\\
&Strict dependency-equivalence (typechecks)& 23.9 & 22.1 & 7.1 & 8.3\\
&Passes \exebench tests &54.4 & 47.4 & -- & -- \\
\hline
\parbox[t]{0mm}{\multirow{4}{*}{\rotatebox[origin=c]{90}{\scriptsize Improvement}}}
&Variable name accuracy & 20.6 & 18.2 & 16.3 & 19.1\\
&Variable type accuracy & 58.2 & 53.2 & 36.4 & 37.9\\
&UDT variable nominal accuracy & 20.7 & 17.2 & 4.0 & 5.7\\
&UDT variable structural accuracy & 34.5 & 44.8 & 9.2 & 15.1\\
\bottomrule
\end{tabular}
\end{subtable}
\vspace{4pt}

\begin{subtable}{1.0\textwidth}
\centering
\caption{CodeLlama-7b\label{tab:codellama_7b}}
\begin{tabular}{clrr|rr}
\toprule
&&exebench&parity-exebench&functions-realtype& \idioms \\
\midrule
\parbox[t]{0mm}{\multirow{4}{*}{\rotatebox[origin=c]{90}{\scriptsize Correctness}}}
&Relaxed dependency-equivalence (consistency)& 33.0 & 29.5 & 26.6 & 27.2\\
&Dependency-equivalence & 32.7 & 29.1 & 17.8 & 18.4\\
&Strict dependency-equivalence (typechecks) & 22.2 & 20.0 & 9.1 & 8.0\\
&Passes \exebench tests & 48.4 & 42.0 & -- & -- \\
\hline
\parbox[t]{0mm}{\multirow{4}{*}{\rotatebox[origin=c]{90}{\scriptsize Improvement}}}
&Variable name accuracy & 19.3 & 18.0 & 17.2 & 19.3\\
&Variable type accuracy & 55.8 & 53.9 & 36.5 & 37.7\\
&UDT variable nominal accuracy & 20.7 & 25.9 & 4.1 & 5.5\\
&UDT variable structural accuracy & 41.4 & 48.1 & 10.0 & 14.1\\
\bottomrule
\end{tabular}
\end{subtable}

\caption{Ablation study. All values are percentages; higher is better. Adjacent columns differ in one experimental condition. \label{fig:main_results}}

\end{table*}

\subsection{RQ3: The Challenge of Real-World UDTs}
\label{sec:rq3-results}

Table~\ref{fig:main_results} shows the results of our ablation study, where we causally show the impact of dataset size, dataset composition, and model context.
The second and third columns (\textbf{parity-exebench} and \textbf{functions-realtype}) show models that are the same in terms of task and context provided, and trained and tested on datasets of similar sizes. However, they differ in dataset complexity, as the \textbf{functions-realtype} configuration trains on \realtype, which has far more, and more complex, UDTs than \exebench (Table~\ref{tab:dataset_complexity}).   

This change causes a substantial decrease in performance on all metrics except variable name accuracy.
For dependency-equivalence (row 2 on each of Tables~\ref{tab:qwen_0.5b}-\ref{tab:codellama_7b}), the drop is 38-42\%, relative to performance on \exebench.
The drop is even steeper when type correctness is factored in (row 3): from 55\%-68\%.
These data highlight the challenge that UDTs provide for real code.
Overall, 
the absence of UDTs in existing benchmarks masks the value of joint code-UDT prediction. 
\realtype helps close the gap.

\begin{tcolorbox}[width=\columnwidth,
                  boxsep=0pt,
                  left=4pt,
                  right=7pt,
                  top=7pt,
                  arc=4pt,
                  boxrule=1pt,
                  toprule=1pt,
                  colback=white
                  ]%
\textbf{Takeaway}: Code with UDTs is substantially more challenging to neurally decompile than code without; strict dependency-equivalence drops by 55\%-68\% with the introduction of more realistic data.
\end{tcolorbox}

\subsection{RQ4: Model Performance and Trends}
\label{sec:rq4-results}

In general, in machine learning, model performance scales with dataset size and parameter counts, but performance gains are usually logarithmic in training set size; doubling either does not lead to a doubling in scores.
We see this as well across all of our, and the results in Table~\ref{fig:main_results}.
For instance, CodeQwen2.5-0.5b, our smallest model, produces dependency-equivalent code 30.7\% on \exebench.
Meanwhile, CodeGemma-7b scores 33.7\%, an increase of about 10\%, despite being 14 times larger.
We also see this in terms of scaling the dataset.
Columns 1 and 2 illustrate this.
Column 1 represents a size increase of over 15 times but the gains in the same metric are only 6-11\%.
The gains on both counts are relatively modest, which may be attributable in the latter case to the fact that the models are pretrained on C code, which means they possess some inherent reasoning ability about C (or, more precisely, they model a good distribution of C code).

The full idioms approach, in the \idioms column, follows a similar trend, albeit on a more difficult dataset (see Section~\ref{sec:rq5}). 
CodeQwen2.5's dependency-equivalence score is 15.6\%, though this drops to 7.2\% when type correctness is also factored in.
Meanwhile, the two 7-billion parameter models, CodeGemma-7b and CodeLlama-7b score 18.3, 8.3 and 18.4, 8.0, respectively.
However, the gains in UDT variable structural accuracy increase more rapidly with model size than other metrics.
(We discuss this rapid gain further in Section~\ref{sec:rq5}.)
The best \idioms model, CodeGemma-7b, recovers structurally accurate UDTs 15.1\% of the time.
To be counted as correct, a prediction must have identical fields, recursively (including field names for nominal struct accuracy).
The UDTs in \realtype are very challenging: the mean recursive number of fields in its UDTs is 17.6 (Table~\ref{tab:dataset_complexity}).

\begin{tcolorbox}[width=\columnwidth,
  boxsep=0pt,
  left=4pt,
  right=7pt,
  top=7pt,
  arc=4pt,
  boxrule=1pt,
  toprule=1pt,
  colback=white
  ]%
\textbf{Takeaway}: Larger models and training sets increase performance modestly across all model types, including \idioms. The best \idioms models achieve dependency-equivalence scores of over 18\% and UDT structural accuracy over 15\% on real code and types.
\end{tcolorbox}

\subsection{RQ5: The Role of Neighboring Context}
\label{sec:rq5}

The last two columns of Table~\ref{fig:main_results} show configurations that control for the utility of the neighboring function content: \textbf{functions-realtype} are models trained and tested on \realtype with only the decompiled function cdoe as context, while \idioms is the full \idioms design including neighboring functions.  
Additional context in the form of neighboring functions' improves performance for many metrics, especially for larger models.
Interestingly, adding context does not have a significant impact on correctness, when measured with dependency-equivalence.
Relaxed and regular dependency-equivalence, have a modest increase or decrease of around 1 (absolute) percentage point.
This is not surprising, because the details necessary to predict code \emph{operations} are all present in the function of interest (except for names).

On the other hand, there is an increase in UDT accuracy that roughly increases with model size, especially in terms of type structure where type and field names are ignored.
As model sizes increase from 0.5b in Table~\ref{tab:qwen_0.5b} to 7b in Table~\ref{tab:codegemma_7b}, gains in UDT composition accuracy increase from 7\% to 11\% to 55\% to 64\%.
CodeLlama (Table~\ref{tab:codellama_7b}), while also a 7b model, scores only a 41\% increase in UDT structural accuracy---there is clearly some variance stemming from the base pretrained model.

Small models also suffer more tradeoffs from attempting to handle the larger context.
CodeGemma-7b and CodeLlama-7b each only suffer a drop in one metric (-3.5\% relaxed dependency-equivalence and -13.8\% in strict dependency-equivalence, respectively), while the smallest model, CodeQwen-0.5b, suffers decreases in 4 of 7 metrics.

\begin{figure}
\begin{lstlisting}[style=cstyle,basicstyle=\ttfamily\bfseries\scriptsize]
struct hash_table {
    int size; 
    struct hash_entry **ht; 
    int (*cmp)(void *, void *); 
    int (*hash)(void *);
};
struct hash_entry {
    void *key;
    void *value;
    struct hash_entry *next; 
};

int hash_table_find(struct hash_table *ht, void *key)
{
    struct hash_entry *he;
    int hash_val = hash_table_hash(ht, key);
    int i = 0;
    for (he = hash_table_get(ht->ht, hash_val); he; 
         he = hash_table_get(ht->ht, hash_val)) {
        if (i++ > ht->size) return -1;
        if (!ht->cmp(he, key)) break;
        hash_val = hash_table_next(ht, hash_val);
    }
    if (he) return hash_val;
    return -1;
}
\end{lstlisting}
\caption{\idioms-functions' prediction when given Figure~\ref{fig:intro_decompiled} as input. The lack of additional context causes a subtle but substantial difference relative to the original source.
The version produced with neighboring context (Figure~\ref{fig:intro_idioms}) is correct.
\label{fig:idioms_functions_prediction}}
\end{figure}

To see the difference that neighboring context can make, compare Figure~\ref{fig:idioms_functions_prediction}, made without additional context, and Figure~\ref{fig:intro_idioms}, made with it.
The function definition in Figure~\ref{fig:idioms_functions_prediction} looks almost correct relative to the original source, when allowances are made for differing but consistent function names.
However, the types are slightly wrong: in addition to missing a current-capacity field, the \cinline{struct} definitions suggest that the table is backed by an array of linked lists---suggesting a separate-chaining collision avoidance strategy, not the robin-hood hashing actually used in the original code.
The identifier names are in turn less reflective of the actual functionality presented in the code.
\idioms-functions interprets the return value as a hash value, not an index into the hash table---likely because indexing into a linked list is uncommon.
Crucially, \emph{the evidence available in the decompiled code (Figure~\ref{fig:intro_decompiled}) is consistent with both \idioms-functions' and \idioms-neighbors predictions}, but only \idioms-functions' prediction is incorrect relative to the original source and misleading.
Details from the additional context help the model be more reflective of the binary.

\begin{tcolorbox}[width=\columnwidth,
                  boxsep=0pt,
                  left=4pt,
                  right=7pt,
                  top=7pt,
                  arc=4pt,
                  boxrule=1pt,
                  toprule=1pt,
                  colback=white
                  ]%
\textbf{Takeaway}: Neighboring context helps increase UDT accuracy with little to no downside in terms of other metrics, especially for larger models.
\end{tcolorbox}

\section{Discussion}
\label{sec:discussion}

Neural decompilation is a difficult task.
\realtype's UDTs average a size of 17.6, making predicting them correctly---even just the types and order of the fields---extremely challenging.
This impacts correctness across multiple metrics; even relaxed dependency equivalence scores are below 30\% on the full \realtype test set. 
However, actual semantic equivalence likely exceeds what these metrics suggest, as functions can be semantically equivalent without structural identity, and the metrics are generally designed to prefer soundness. 
The true rate of semantic equivalence is likely higher than these metrics suggest.
That said, \idioms significantly outperforms prior work despite these challenges, using both conservative sound metrics like structural equivalence, or the \exebench test cases that evaluate partial correctness. 

Code that is structured in the same way as the original (something these metrics implicitly measure) is a positive sign about model performance in general, and likely useful in practice.
Indeed, code \emph{improvement} in the form of better naming or structure is a key potential benefit, and \idioms also achieves state-of-the-art performance here. 
Name prediction accuracy is limited in part by the metric's strict matching requirements.  Note that many acceptable name variations exist (e.g., \cinline{length} vs. \cinline{len} vs. \cinline{size}),  which name accuracy will not recognize.  Relaxing metrics to allow common aliases could yield more optimistic results.
Meanwhile, type accuracy scores (30-40\% for \idioms) at least partially reflect inherent ambiguities in primitive type sizes, as well as the prevalence of UDTs (26.4\% of \realtype variables). Indeed, the latter fact explains most of the type accuracy performance differences observed between the \realtype and \exebench evaluaions.

The source of difficult in neurally decompiling real code can often be traced back to UDTs.
However, fundamentally, it is also difficult to \emph{evaluate} acceptability, which boils down to questions like semantic equivalence or variable name quality.  Our results should further motivate ongoing efforts to iterate on good design, datasets, and metrics for improvements in neural decompilation. 

\vspace{0.5ex}
\noindent\textbf{Limitations and Threats.}
A key threat to validity with most work involving large language models is data leakage through pretraining.
Most LLMs include code from GitHub, from which we also draw \realtype.
Despite this, we think the risk of data leakage is small.
Relatively little decompiled code is found on GitHub or on the Internet in general, never mind the crucial decompiled-to-original mapping needed for neural decompilation; models are likely unfamiliar with it.

Our dataset is constructed of open source projects on GitHub, that compile.
It is possible that some decompilation targets, especially malware, may have systematic differences from our data and thus affect performance.
Malware in particular is 
often \emph{obfuscated}, or transformed in a way that makes them more difficult to understand.
We view deobfuscation as an orthogonal problem. There are a variety of deobfuscation techniques~\cite{dong2022cadecff,coogan2011deobfuscation,david2020qsynth,liang2018deobfuscation,tofighi2018dose,you2022deoptfuscator} that can be applied to de-obfuscate code before neurally decompiling it. 
However, in cases where these fail or a novel obfuscation is encountered it may be necessary to input the obfuscated code to the neural decompiler directly.
We leave studying the impacts of this to future work.

\section{Related work}

\noindent\textbf{Neural Decompilation.}
Early neural decompilation evolved from RNNs~{katz2018} through various architectures~\cite{coda,cao2022boosting}, to 
transformers~\cite{hosseinibeyond}.
The dominance of the transformer architecture~\cite{transformer} has driven more recent work to leverage its potential: SLaDe~\cite{slade} outperforms prior work, but requires test cases for inference, which is unrealistic in most reverse engineering scenarios.  
LLM4Decompile~\cite{llm4decompile} introduce a family of large causal transformer models with sizes in the billions of parameters, accepting either assembly or Ghidra output as input. 
Nova~\cite{nova}, which operates directly on assembly, uses hierarchical attention to adapt to and handle typically long sequences of assembly instructions.
We compare to these latter two as baselines in our evaluation.
Note that general-purpose LLMs like ChatGPT perform very poorly at decompilation~\cite{slade,llm4decompile,nova}, likely because their training data contains little decompiled code. 

Unlike prior work, we recast neural decompilation as joint code and type definition recovery, and are the first to leverage interprocedural context.  We demonstrate state-of-the-art performance across a variety of metrics.  Importantly, our results also demonstrate the importance of evaluating neural decompilers on realistic code with complex, user-defined types, and contribute a novel dataset to help close this gap.

\vspace{0.5ex}
\noindent\textbf{Type Recovery.}
Approaches specific to type recovery in decompiled code range from constraint-based~\cite{tie} to probabilistic methods.  Dynamic approaches like REWARD~\cite{reward} and Howard~\cite{howard} require execution traces, which are not always available in reverse engineering context.  
 There is sufficient information available in an executable to deterministically eliminate some types, but not enough to full defnitions, nor reconstruct meaningful names. Fundamentally, all non-probabilistic techniques are fundamentally limited by the information present in the binary.

Relevant ML-based approaches include Osprey~\cite{osprey} and TyGr~\cite{tygr}, for structural type prediction (not type names), and DIRTY~\cite{dirty} for joint name-type recovery based on a fixed library.  
ReSym~\cite{resym} extends this idea, with arbitrary field name generation. 

These prior approaches either do not fully reconstruct \emph{and} name UDTs, cannot handle nested structures, or are limited to predicting names and layouts drawn from predefined type libraries. In contrast, \idioms is capable of generating complete, arbitrary type definitions including such nested structures.
To be counted as correct in our experiments, type definitions for all nested structures must have equivalent composition; nested structures are extremely common in our real-world data.

\section{Conclusion}

In this work, we introduce the \idioms family of neural decompilers, which perform joint code and type definition recovery.
We build a dataset, \realtype, which contains 157,163 total functions and the definitions of their user-defined types.
We train models on \exebench and \realtype, and show that \realtype is a more difficult dataset despite their functions being of a similar size.
Additionally, we show that additional neighboring context helps address the scattered evidence problem and thus improves performance on UDTs.
We've illustrated that joint code-type definition recovery is crucial but a difficult problem in practice.
We hope this motivates future work in tackling a key challenge in tackling neural decompilation.

\section*{Acknowledgment}

This material is based upon work supported in part by the National Science Foundation (award DGE2140739).

\bibliographystyle{IEEEtran}
\bibliography{references}

\begin{thebibliography}{10}
\providecommand{\url}[1]{#1}
\csname url@samestyle\endcsname
\providecommand{\newblock}{\relax}
\providecommand{\bibinfo}[2]{#2}
\providecommand{\BIBentrySTDinterwordspacing}{\spaceskip=0pt\relax}
\providecommand{\BIBentryALTinterwordstretchfactor}{4}
\providecommand{\BIBentryALTinterwordspacing}{\spaceskip=\fontdimen2\font plus
\BIBentryALTinterwordstretchfactor\fontdimen3\font minus
  \fontdimen4\font\relax}
\providecommand{\BIBforeignlanguage}[2]{{%
\expandafter\ifx\csname l@#1\endcsname\relax
\typeout{** WARNING: IEEEtran.bst: No hyphenation pattern has been}%
\typeout{** loaded for the language `#1'. Using the pattern for}%
\typeout{** the default language instead.}%
\else
\language=\csname l@#1\endcsname
\fi
#2}}
\providecommand{\BIBdecl}{\relax}
\BIBdecl

\bibitem{dire}
J.~Lacomis, P.~Yin, E.~Schwartz, M.~Allamanis, C.~Le~Goues, G.~Neubig, and
  B.~Vasilescu, ``Dire: A neural approach to decompiled identifier naming,'' in
  \emph{2019 34th IEEE/ACM International Conference on Automated Software
  Engineering (ASE)}.\hskip 1em plus 0.5em minus 0.4em\relax IEEE, 2019, pp.
  628--639.

\bibitem{direct}
V.~Nitin, A.~Saieva, B.~Ray, and G.~Kaiser, ``Direct: A transformer-based model
  for decompiled variable name recovery,'' \emph{NLP4Prog 2021}, p.~48, 2021.

\bibitem{varbert}
K.~K. Pal, A.~P. Bajaj, P.~Banerjee, A.~Dutcher, M.~Nakamura, Z.~L. Basque,
  H.~Gupta, S.~A. Sawant, U.~Anantheswaran, Y.~Shoshitaishvili \emph{et~al.},
  ``" len or index or count, anything but v1": Predicting variable names in
  decompilation output with transfer learning,'' in \emph{2024 IEEE Symposium
  on Security and Privacy (SP)}.\hskip 1em plus 0.5em minus 0.4em\relax IEEE
  Computer Society, 2024, pp. 152--152.

\bibitem{nero}
Y.~David, U.~Alon, and E.~Yahav, ``Neural reverse engineering of stripped
  binaries using augmented control flow graphs,'' \emph{Proceedings of the ACM
  on Programming Languages}, vol.~4, no. OOPSLA, pp. 1--28, 2020.

\bibitem{symlm}
X.~Jin, K.~Pei, J.~Y. Won, and Z.~Lin, ``Symlm: Predicting function names in
  stripped binaries via context-sensitive execution-aware code embeddings,'' in
  \emph{Proceedings of the 2022 ACM SIGSAC Conference on Computer and
  Communications Security}, 2022, pp. 1631--1645.

\bibitem{kim2023transformer}
H.~Kim, J.~Bak, K.~Cho, and H.~Koo, ``A transformer-based function symbol name
  inference model from an assembly language for binary reversing,'' in
  \emph{Proceedings of the 2023 ACM Asia Conference on Computer and
  Communications Security}, 2023, pp. 951--965.

\bibitem{lehmann2022finding}
D.~Lehmann and M.~Pradel, ``Finding the dwarf: recovering precise types from
  webassembly binaries,'' in \emph{Proceedings of the 43rd ACM SIGPLAN
  International Conference on Programming Language Design and Implementation},
  2022, pp. 410--425.

\bibitem{osprey}
Z.~Zhang, Y.~Ye, W.~You, G.~Tao, W.-c. Lee, Y.~Kwon, Y.~Aafer, and X.~Zhang,
  ``Osprey: Recovery of variable and data structure via probabilistic analysis
  for stripped binary,'' in \emph{2021 IEEE Symposium on Security and Privacy
  (SP)}.\hskip 1em plus 0.5em minus 0.4em\relax IEEE, 2021, pp. 813--832.

\bibitem{tygr}
C.~Zhu, Z.~Li, A.~Xue, A.~P. Bajaj, W.~Gibbs, Y.~Liu, R.~Alur, T.~Bao, H.~Dai,
  A.~Doup{\'e} \emph{et~al.}, ``$\{$TYGR$\}$: Type inference on stripped
  binaries using graph neural networks,'' in \emph{33rd USENIX Security
  Symposium (USENIX Security 24)}, 2024, pp. 4283--4300.

\bibitem{dirty}
Q.~Chen, J.~Lacomis, E.~J. Schwartz, C.~Le~Goues, G.~Neubig, and B.~Vasilescu,
  ``Augmenting decompiler output with learned variable names and types,'' in
  \emph{31st USENIX Security Symposium (USENIX Security 22)}, 2022, pp.
  4327--4343.

\bibitem{hext5}
J.~Xiong, G.~Chen, K.~Chen, H.~Gao, S.~Cheng, and W.~Zhang, ``Hext5: Unified
  pre-training for stripped binary code information inference,'' in \emph{2023
  38th IEEE/ACM International Conference on Automated Software Engineering
  (ASE)}.\hskip 1em plus 0.5em minus 0.4em\relax IEEE, 2023, pp. 774--786.

\bibitem{resym}
D.~Xie, Z.~Zhang, N.~Jiang, X.~Xu, L.~Tan, and X.~Zhang, ``Resym: Harnessing
  llms to recover variable and data structure symbols from stripped binaries,''
  in \emph{Proceedings of the 2024 on ACM SIGSAC Conference on Computer and
  Communications Security}, 2024, pp. 4554--4568.

\bibitem{dramko2024taxonomy}
L.~Dramko, J.~Lacomis, E.~J. Schwartz, B.~Vasilescu, and C.~Le~Goues, ``A
  taxonomy of c decompiler fidelity issues,'' in \emph{33th USENIX Security
  Symposium (USENIX Security 24)}, 2024.

\bibitem{nova}
N.~Jiang, C.~Wang, K.~Liu, X.~Xu, L.~Tan, and X.~Zhang, ``Nova+: Generative
  language models for binaries,'' \emph{arXiv preprint arXiv:2311.13721}, 2023.

\bibitem{degpt}
P.~Hu, R.~Liang, and K.~Chen, ``Degpt: Optimizing decompiler output with llm,''
  2024.

\bibitem{llm4decompile}
\BIBentryALTinterwordspacing
H.~Tan, Q.~Luo, J.~Li, and Y.~Zhang, ``{LLM}4{D}ecompile: Decompiling binary
  code with large language models,'' in \emph{Proceedings of the 2024
  Conference on Empirical Methods in Natural Language Processing},
  Y.~Al-Onaizan, M.~Bansal, and Y.-N. Chen, Eds.\hskip 1em plus 0.5em minus
  0.4em\relax Miami, Florida, USA: Association for Computational Linguistics,
  Nov. 2024, pp. 3473--3487. [Online]. Available:
  \url{https://aclanthology.org/2024.emnlp-main.203/}
\BIBentrySTDinterwordspacing

\bibitem{votipka2020observational}
D.~Votipka, S.~Rabin, K.~Micinski, J.~S. Foster, and M.~L. Mazurek, ``An
  observational investigation of reverse engineers’ processes,'' in
  \emph{29th USENIX Security Symposium (USENIX Security 20)}, 2020, pp.
  1875--1892.

\bibitem{tiesurvey}
\BIBentryALTinterwordspacing
J.~Caballero and Z.~Lin, ``Type inference on executables,'' \emph{ACM Comput.
  Surv.}, vol.~48, no.~4, May 2016. [Online]. Available:
  \url{https://doi.org/10.1145/2896499}
\BIBentrySTDinterwordspacing

\bibitem{slade}
J.~Armengol-Estapé, J.~Woodruff, C.~Cummins, and M.~F. O'Boyle, ``Slade: A
  portable small language model decompiler for optimized assembly,'' in
  \emph{2024 IEEE/ACM International Symposium on Code Generation and
  Optimization (CGO)}, 2024, pp. 67--80.

\bibitem{exebench}
\BIBentryALTinterwordspacing
J.~Armengol-Estap\'{e}, J.~Woodruff, A.~Brauckmann, J.~W. d.~S. Magalh\~{a}es,
  and M.~F.~P. O'Boyle, ``Exebench: an ml-scale dataset of executable c
  functions,'' in \emph{Proceedings of the 6th ACM SIGPLAN International
  Symposium on Machine Programming}, ser. MAPS 2022.\hskip 1em plus 0.5em minus
  0.4em\relax New York, NY, USA: Association for Computing Machinery, 2022, p.
  50–59. [Online]. Available: \url{https://doi.org/10.1145/3520312.3534867}
\BIBentrySTDinterwordspacing

\bibitem{katz2018}
D.~S. Katz, J.~Ruchti, and E.~Schulte, ``Using recurrent neural networks for
  decompilation,'' in \emph{IEEE International Conference on Software Analysis,
  Evolution and Reengineering}, 4 2018, pp. 346--356.

\bibitem{coda}
C.~Fu, H.~Chen, H.~Liu, X.~Chen, Y.~Tian, F.~Koushanfar, and J.~Zhao, ``Coda:
  An end-to-end neural program decompiler,'' \emph{Advances in Neural
  Information Processing Systems}, vol.~32, 2019.

\bibitem{cao2022boosting}
Y.~Cao, R.~Liang, K.~Chen, and P.~Hu, ``Boosting neural networks to decompile
  optimized binaries,'' in \emph{Proceedings of the 38th Annual Computer
  Security Applications Conference}, 2022, pp. 508--518.

\bibitem{hosseinibeyond}
I.~Hosseini and B.~Dolan-Gavitt, ``Beyond the c: Retargetable decompilation
  using neural machine translation,'' 2022.

\bibitem{ghcc}
\BIBentryALTinterwordspacing
Z.~Hu. (2021) Ghcc. [Online]. Available: \url{https://github.com/huzecong/ghcc}
\BIBentrySTDinterwordspacing

\bibitem{spinellis2020dataset}
D.~Spinellis, Z.~Kotti, and A.~Mockus, ``A dataset for github repository
  deduplication,'' in \emph{Proceedings of the 17th international conference on
  mining software repositories}, 2020, pp. 523--527.

\bibitem{minhashing}
A.~Z. Broder, ``On the resemblance and containment of documents,'' in
  \emph{Proceedings. Compression and Complexity of SEQUENCES 1997 (Cat. No.
  97TB100171)}.\hskip 1em plus 0.5em minus 0.4em\relax IEEE, 1997, pp. 21--29.

\bibitem{TheStack}
D.~Kocetkov, R.~Li, L.~Ben~Allal, J.~Li, C.~Mou, C.~Muñoz~Ferrandis,
  Y.~Jernite, M.~Mitchell, S.~Hughes, T.~Wolf, D.~Bahdanau, L.~von Werra, and
  H.~de~Vries, ``The stack: 3 tb of permissively licensed source code,''
  \emph{Preprint}, 2022.

\bibitem{ThePile}
L.~Gao, S.~Biderman, S.~Black, L.~Golding, T.~Hoppe, C.~Foster, J.~Phang,
  H.~He, A.~Thite, N.~Nabeshima \emph{et~al.}, ``The pile: An 800gb dataset of
  diverse text for language modeling,'' \emph{arXiv preprint arXiv:2101.00027},
  2020.

\bibitem{holtzman2019curious}
A.~Holtzman, J.~Buys, L.~Du, M.~Forbes, and Y.~Choi, ``The curious case of
  neural text degeneration,'' \emph{arXiv preprint arXiv:1904.09751}, 2019.

\bibitem{codealign}
L.~Dramko, C.~{Le Goues}, and E.~J. Schwartz, ``Fast, fine-grained equivalence
  checking for neural decompilers,'' \emph{arXiv preprint arXiv:2501.04811},
  2025.

\bibitem{yang1989detecting}
W.~Yang, S.~Horwitz, and T.~Reps, ``Detecting program components with
  equivalent behaviors,'' University of Wisconsin-Madison Department of
  Computer Sciences, Tech. Rep., 1989.

\bibitem{codegemma}
C.~Team, H.~Zhao, J.~Hui, J.~Howland, N.~Nguyen, S.~Zuo, A.~Hu, C.~A.
  Choquette-Choo, J.~Shen, J.~Kelley \emph{et~al.}, ``Codegemma: Open code
  models based on gemma,'' \emph{arXiv preprint arXiv:2406.11409}, 2024.

\bibitem{qlora}
T.~Dettmers, A.~Pagnoni, A.~Holtzman, and L.~Zettlemoyer, ``Qlora: Efficient
  finetuning of quantized llms,'' \emph{Advances in Neural Information
  Processing Systems}, vol.~36, 2024.

\bibitem{qwen}
B.~Hui, J.~Yang, Z.~Cui, J.~Yang, D.~Liu, L.~Zhang, T.~Liu, J.~Zhang, B.~Yu,
  K.~Lu \emph{et~al.}, ``Qwen2. 5-coder technical report,'' \emph{arXiv
  preprint arXiv:2409.12186}, 2024.

\bibitem{codellama}
B.~Roziere, J.~Gehring, F.~Gloeckle, S.~Sootla, I.~Gat, X.~E. Tan, Y.~Adi,
  J.~Liu, R.~Sauvestre, T.~Remez \emph{et~al.}, ``Code llama: Open foundation
  models for code,'' \emph{arXiv preprint arXiv:2308.12950}, 2023.

\bibitem{jin2023binary}
X.~Jin, J.~Larson, W.~Yang, and Z.~Lin, ``Binary code summarization:
  Benchmarking chatgpt/gpt-4 and other large language models,'' \emph{arXiv
  preprint arXiv:2312.09601}, 2023.

\bibitem{dong2022cadecff}
W.~Dong, J.~Lin, R.~Chang, and R.~Wang, ``Cadecff: Compiler-agnostic
  deobfuscator of control flow flattening,'' in \emph{Proceedings of the 13th
  Asia-Pacific Symposium on Internetware}, 2022, pp. 282--291.

\bibitem{coogan2011deobfuscation}
K.~Coogan, G.~Lu, and S.~Debray, ``Deobfuscation of virtualization-obfuscated
  software: a semantics-based approach,'' in \emph{Proceedings of the 18th ACM
  conference on Computer and communications security}, 2011, pp. 275--284.

\bibitem{david2020qsynth}
R.~David, L.~Coniglio, M.~Ceccato \emph{et~al.}, ``Qsynth-a program synthesis
  based approach for binary code deobfuscation,'' in \emph{BAR 2020 Workshop},
  2020.

\bibitem{liang2018deobfuscation}
M.~Liang, Z.~Li, Q.~Zeng, and Z.~Fang, ``Deobfuscation of
  virtualization-obfuscated code through symbolic execution and compilation
  optimization,'' in \emph{Information and Communications Security: 19th
  International Conference, ICICS 2017, Beijing, China, December 6-8, 2017,
  Proceedings 19}.\hskip 1em plus 0.5em minus 0.4em\relax Springer, 2018, pp.
  313--324.

\bibitem{tofighi2018dose}
R.~Tofighi-Shirazi, M.~Christofi, P.~Elbaz-Vincent, and T.-H. Le, ``Dose:
  Deobfuscation based on semantic equivalence,'' in \emph{Proceedings of the
  8th Software Security, Protection, and Reverse Engineering Workshop}, 2018,
  pp. 1--12.

\bibitem{you2022deoptfuscator}
G.~You, G.~Kim, S.~Han, M.~Park, and S.-J. Cho, ``Deoptfuscator: Defeating
  advanced control-flow obfuscation using android runtime (art),'' \emph{IEEE
  Access}, vol.~10, pp. 61\,426--61\,440, 2022.

\bibitem{transformer}
\BIBentryALTinterwordspacing
A.~Vaswani, N.~Shazeer, N.~Parmar, J.~Uszkoreit, L.~Jones, A.~N. Gomez, L.~u.
  Kaiser, and I.~Polosukhin, ``Attention is all you need,'' in \emph{Advances
  in Neural Information Processing Systems}, I.~Guyon, U.~V. Luxburg,
  S.~Bengio, H.~Wallach, R.~Fergus, S.~Vishwanathan, and R.~Garnett, Eds.,
  vol.~30.\hskip 1em plus 0.5em minus 0.4em\relax Curran Associates, Inc.,
  2017. [Online]. Available:
  \url{https://proceedings.neurips.cc/paper_files/paper/2017/file/3f5ee243547dee91fbd053c1c4a845aa-Paper.pdf}
\BIBentrySTDinterwordspacing

\bibitem{tie}
J.~Lee, T.~Avgerinos, and D.~Brumley, ``Tie: Principled reverse engineering of
  types in binary programs,'' 2011.

\bibitem{reward}
Z.~Lin, X.~Zhang, and D.~Xu, ``Automatic reverse engineering of data structures
  from binary execution,'' in \emph{Proceedings of the 11th Annual Information
  Security Symposium}, 2010.

\bibitem{howard}
A.~Slowinska, T.~Stancescu, and H.~Bos, ``Howard: A dynamic excavator for
  reverse engineering data structures.'' in \emph{Network and Distributed
  System Security Symposium (NDSS)}, 2011.

\bibitem{loshchilov2016sgdr}
I.~Loshchilov and F.~Hutter, ``Sgdr: Stochastic gradient descent with warm
  restarts,'' \emph{arXiv preprint arXiv:1608.03983}, 2016.

\end{thebibliography}

\section*{Appendix}

We used the following conventions for training models in our experiments.
When training on the full compilable partition of the \exebench dataset (for comparison with \realtype), we trained CodeQwen for 8 epochs, the 1-2 billion parameter models for two epochs, and the largest models for 1 epoch.
We find that the larger models need less finetuning before they converge; they are inherently more capable.
Simultaneously, larger models are more expensive to finetune.
Uniquely among the models we use, LLM4Decompile has been pretrained on decompiled code, though it was trained on code decompiled with Ghidra rather than Hex-Rays.
We train all models with a cosine learning rate scheduler~\cite{loshchilov2016sgdr} starting from a learning rate of $5\times 10^{-5}$.
We use a batch size of 64.

For our function-context models, we configure the context window---the maximum amount of tokens the model will process at once---to be up to 2048 tokens, 1024 of which is reserved for the original code tokens and UDT definitions.
We configure the context window size for \idioms models with neighboring context to be up to 4096 tokens, 3072 of which is the decompiled function and neighboring context and 1024 of which is reserved for the original code and UDTs.

\end{document}